\documentclass[12pt]{iopart}

\usepackage{booktabs}
\usepackage{graphicx}
\usepackage{subfigure}
\usepackage{indentfirst}
\usepackage{bm}
\usepackage{multirow}
\usepackage{color}
\expandafter\let\csname equation*\endcsname\relax
\expandafter\let\csname endequation*\endcsname\relax
\usepackage{amsmath}
\graphicspath{{./}}

\begin{document}

\title[]{The single-particle space-momentum angle distribution effect on two-pion HBT correlation with the Coulomb interaction}

\author{Hang Yang, Qichun Feng, Jingbo Zhang*}

\address{School of Physics, Harbin Institute of Technology, Harbin 150001, China}

\ead{jinux@hit.edu.cn}
\vspace{10pt}

\begin{abstract}
We calculate the HBT radius $R_{\rm s}$ for $\pi^+$ with the Coulomb interaction by using the string melting version of a multiphase transport(AMPT) model. We study the relationship between the single-particle space-momentum angle and the particle sources and discuss HBT radii without single-particle space-momentum correlation. Additionally, we study the Coulomb interaction effect on the numerical connection between the single-particle space-momentum angle distribution and the transverse momentum dependence of $R_{\rm s}$.
\end{abstract}
\vspace{2pc}
\maketitle
%%%%%%%%%%%%%%%%%%%%%%%%%%%%%%%%%%%%%%%%%%%%%%%%%%%%%%%%%%%%%%%%%%%%%%%%%%%%%%%%%%%%%%%%%%%%%%
\section{Introduction}
The Hanbury-Brown Twiss (HBT) method is a useful tool in relativistic heavy ion collisions, and it can probe the dynamically generated geometry structure of the emitting system. It is also named the two-pion interferometry method, for it is often used with the pion which are the most particles in high-energy collisions. The method of measuring the photons correlation to extract angular sizes of stars was invented by Hanbury Brown and Twiss in astronomy in the 1950s\cite{HanburyBrown:1956bqd}. Several years later, this method was extended in $\overline{p}+p$ collisions by G. Goldhaber, S. Goldhaber, W. Lee and A. Pais\cite{PhysRev.120.300}. After years of improvement, the HBT method has become a precision tool for measuring space-time and dynamic properties of the emitting source\cite{doi:10.1146/annurev.nucl.55.090704.151533,doi:10.1146/annurev.nucl.49.1.529,WIEDEMANN1999145}, and it has been used in $e^{+}+e^{-}$, hadron+hadron, and heavy ion collisions\cite{2001AcPPB,2003RPPh,2005ARNPS}.

At the high energies of the heavy-ion collisions,{the normal matter transforms into the Quark-Gluon Plasma (QGP), and it is a new state of matter consisting of deconfined quarks and gluons\cite{ADAMS2005102,ADCOX2005184}. There are two kinds of phase transition between the low-temperature hadronic phase and the high-temperature quark-gluon plasma phase, cross-over transition, and first-order transition. And the critical end point(CEP) is the point where the first-order phase transition terminates\cite{Borsanyi:2010bp,PhysRevD.67.014028}. Searching for the CEP at lower energies on the QCD phase diagram is one main goal of the Beam Energy Scan(BES) program\cite{PhysRevLett.81.4816,2010arXiv1007.2613S}. There will be critical behavior near the CEP\cite{doi:10.1142/S0217751X05027965}, and the transport coefficients will change violently, which will lead the changes in HBT radii, then the CEP can be estimated by the HBT analysis\cite{PhysRevLett.114.142301}.

The space-momentum correlation is important in HBT research, which is caused by the collective expanding behavior of the collision source\cite{PhysRevLett.53.1219}. Moreover, the space-momentum correlation can lead to the changing of the HBT radii with the transverse momentum of the pion pairs\cite{PhysRevC.53.918}. This phenomenon is named the transverse momentum dependence of HBT radii. And our research is focusing on the connection between the space-momentum correlation and this phenomenon. For we need a tool to quantify this space-momentum correlation, the normalized single-particle space-momentum angle distribution has been introduced in our previous work\cite{Yang_2020,Yang_2021}. This distribution consists of a series of angles belonging to freeze-out pions in the same energy sections and the same transverse momentum pion pair sections. We use the projection angle $\Delta \theta$ on the transverse plane in our study, as shown in Figure~\ref{fig_pr}. With this angle distribution, we can obtain more information about the source from the transverse momentum dependence of HBT radii.
\begin{figure}[htb]
	\centering
	\includegraphics[scale=0.13]{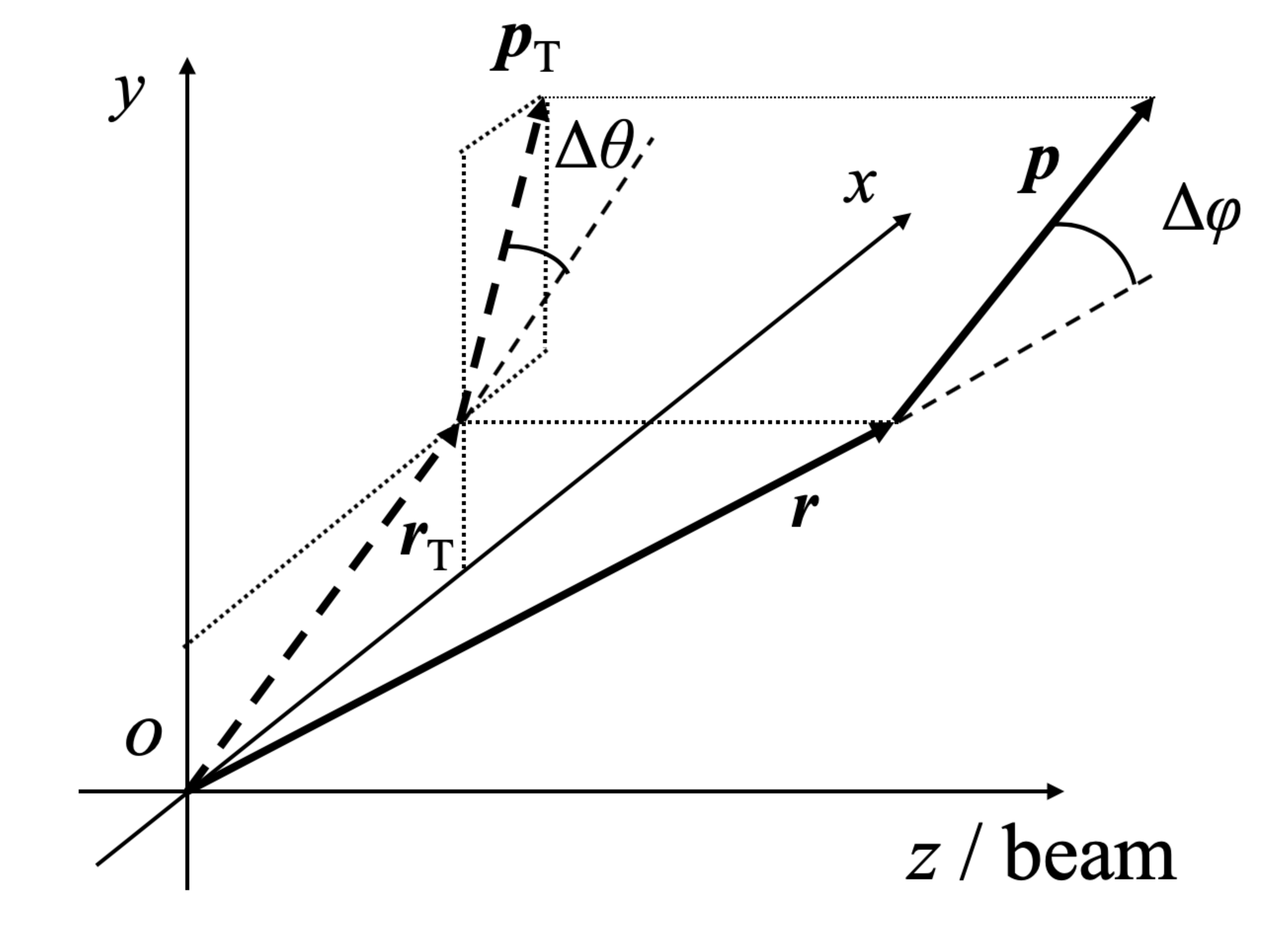}
	\caption{The diagram of the $\Delta\varphi$ and $\Delta \theta$. $\Delta\varphi$ is the angle between $\bm r$ and $\bm p$, and $\Delta \theta$ is the angle between $\bm r_{\rm T}$ and $\bm p_{\rm T}$, at the freeze-out time. The origin is the center of the source.}
	\label{fig_pr}
\end{figure}

In our previous work, we treated three kinds of pions as one, considered them to have the same mass, and neglected the Coulomb interaction\cite{Yang_2020,Yang_2021}. While in experiments, the three kinds of pions can be distinguished and the charged ones are more easily to be detected. Therefore, in this paper, we only focus on the $\pi^+$ and discuss the influence of the Coulomb interaction on the numerical connection between the single-particle space-momentum angle $\Delta \theta$ distribution and the transverse momentum dependence of HBT radius $R_{\rm s}$. We use a multiphase transport (AMPT) model to generate the freeze-out $\pi^+$ at different collision energies. This model contains the physical processes of the relativistic heavy-ion collisions. Besides, it has already been widely used in HBT research\cite{Shan_2009,PhysRevC.92.014909,PhysRevC.96.044914}.

The paper is organized as follows. In Sec. II, the string melting AMPT model and the HBT correlation are briefly introduced. In Sec. III, the HBT radius $R_{\rm s}$ for $\pi^+$ are calculated in two conditions, with and without the Coulomb interaction. In Sec. IV, we discuss the influence of the single-particle space-momentum angle $\Delta \theta$ distribution on the HBT radius $R_{\rm s}$, and we build the numerical connections between them with and without the Coulomb interaction. In the final section, we give the summary. 
%%%%%%%%%%%%%%%%%%%%%%%%%%%%%%%%%%%%%%%%%%%%%%%%%%%%%%%%%%%%%%%%%%%%%%%%%%%%%%%%%%%%%%%%%%%
\section{The string melting AMPT model and methodology }

In this paper, we use the string melting AMPT model, which can give a better description of the correlation function in HBT research\cite{PhysRevC.72.064901,PhysRevLett.89.152301}. The string melting AMPT model contains four main parts. The first part uses the HIJING model to produce the partons and strings, and the strings will fragment into partons. The second part uses Zhang’s parton-cascade(ZPC) model to describe the interactions among these partons. Then in the third part, a quark coalescence model is used to combine these partons into hadrons. In the last part, these hadrons interaction is described by a relativistic transport(ART) model till hadrons freeze out. With the models in these four main parts, the string melting AMPT model can be used to describe the heavy-ion collisions.

The HBT radii are important in HBT research, they can be extracted by the HBT three-dimensional correlation function, and the normal form can be written as\cite{Alexander_2003}
\begin{equation}
\label{fit_function1}
	C(\bm{q},\bm{K})=1 + \lambda{e}^
		{-q_{\rm o}^2 R_{\rm o}^2(\bm K)-q_{\rm s}^2 R_{\rm s}^2(\bm K)-
		q_{\rm l}^2 R_{\rm l}^2(\bm K)-2q_{\rm o}q_{\rm l} R_{\rm ol}^2(\bm K)}.
\end{equation}
In addition, in this paper, we also use the form with the Coulomb interaction as\cite{PhysRevC.71.044906}
\begin{equation}
\label{fit_function2}
	C(\bm{q},\bm{K})=1-\lambda + \lambda K_{coul}\big[ 1+ {e}^
		{-q_{\rm o}^2 R_{\rm o}^2(\bm K)-q_{\rm s}^2 R_{\rm s}^2(\bm K)-
		q_{\rm l}^2 R_{\rm l}^2(\bm K)-2q_{\rm o}q_{\rm l} R_{\rm ol}^2(\bm K)}\big],
\end{equation}
where ${\bm q}={\bm p}_1-{\bm p}_2$, ${\bm K}=({\bm p}_1+{\bm p}_2)/2$, $C$ is the two pion correlation function, $K_{coul}$ is the squared Coulomb wave function, and the $\lambda$ is the coherence parameter. $R$ is for the HBT radius. It is usually used in the `out-side-long' coordinate system, and this system is used for the pair particles, as shown in Figure~\ref{fig_o-s-l}. The l is the longitudinal direction, and also the beam direction. The o and s are outward and sideward directions that are both defined on the transverse plane. The pair particles momentum direction is defined as the outward direction, then the side direction is the direction perpendicular to the outward direction. The $R_{\rm ol}$ is the cross term, it will vanish at mid-rapidity in a symmetric system. In this paper, we set the biggest impact parameter as 1.4 fm, and chose the mid-rapidity range $-0.5<\eta<0.5$. We use Equation~(\ref{fit_function1}) to fit the correlation function without the Coulomb interaction and use Equation~(\ref{fit_function2}) to fit the correlation function with the Coulomb interaction.

\begin{figure}[htb]
	\centering
	\includegraphics[scale=0.3]{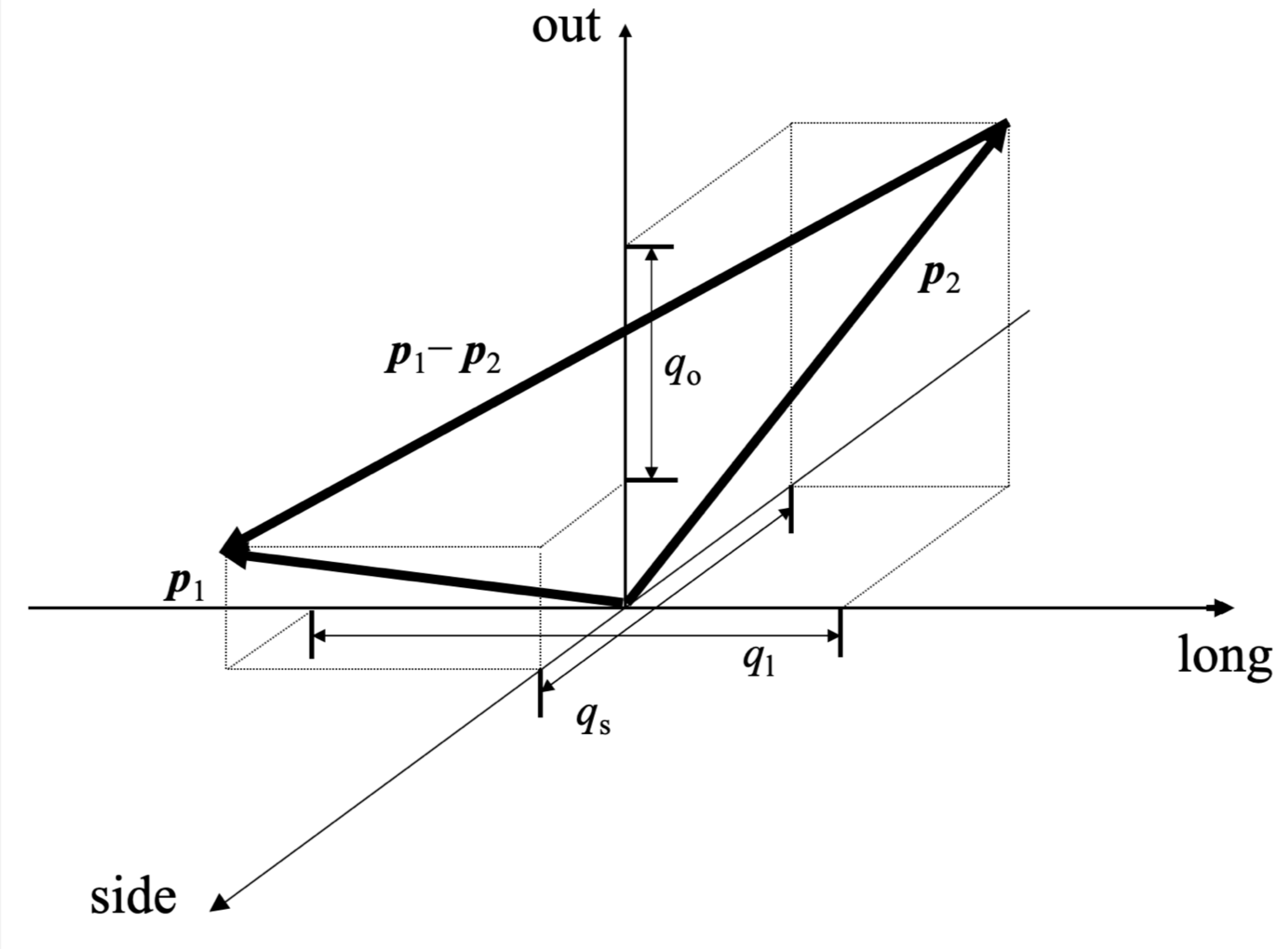}
	\caption{The diagram of `out-side-long'(o-s-l) coordinate system.}
	\label{fig_o-s-l}
\end{figure} 

We use the Correlation After Burner (CRAB) code to generate the correlation function, it can read the phase space information from the string melting AMPT model\cite{CRAB:2006}. It generates the correlation function by the formula
\begin{equation}\label{c2}
	C(\bm{q},\bm{K})=1+\frac{\int {\rm d}^4x_1 {\rm d}^4x_2 S_1(x_1,{\bm p}_2) S_2(x_2,{\bm p}_2)
		{\left|\psi_{\rm{rel}} \right|}^{2}}
		{\int {\rm d}^{4}x_{1}{\rm d}^{4}x_{2}S_{1}(x_{1},{\bm p}_{2})S_{2}(x_{2},{\bm p}_{2})},
\end{equation}
where $\psi_{\rm{rel}}$ is the relative two particles wave function which includes the interaction between two particles. $S(x,{\bm p})$ is the single particle emission function. 
%%%%%%%%%%%%%%%%%%%%%%%%%%%%%%%%%%%%%%%%%%%%%%%%%%%%%%%%%%%%%%%%%%%%%%%%%%%%%%%%%%%%%%%%%%%%
\section{The Coulomb effect on $R_{\rm s}$}

We produced the particles of Au+Au collisions at $\sqrt{S_{NN}}=$14.5, 19.6, 27, 39, 62.4, 200 GeV by the string melting AMPT model. These energies are the BES energies, each energy has more than sixty thousand collision events. We separate the pair transverse momentum 125--625 MeV/c into 9 bins. While the last bin has fewer particles, we set it for 525-625 MeV/c, and the width is twice that of the others. Then we use the CRAB code to calculate the correlation functions for the $\pi^+$. The Coulomb interaction can change the correlation functions, which are shown in Figure~\ref{fit_c}. If neglect the Coulomb interaction, because of the collective flow created by the collisions, two $\pi^+$ are easily frozen out along the same direction, and the correlation function will be a Gaussian form. While considering the Coulomb interaction, the two $\pi^+$ have the Coulomb potential between them, and it will change the two $\pi^+$ wave function $\psi_{\rm{rel}}$. The closer of the two $\pi^+$, the bigger the Coulomb potential between them. When two $\pi^+$ are close to each other, the probability density of the two $\pi^+$ will decrease, and it will cause the decrease of the Correlation function in the direction along the momentum difference direction. $q_{\rm s}$ is related to the momentum difference on the transverse plane, so there is a little gap in the correlation function in the $q_{\rm s}$ direction.

\begin{figure}[!htb]
	\centering
	\subfigure[]
	{
		\includegraphics[width=0.47\textwidth]{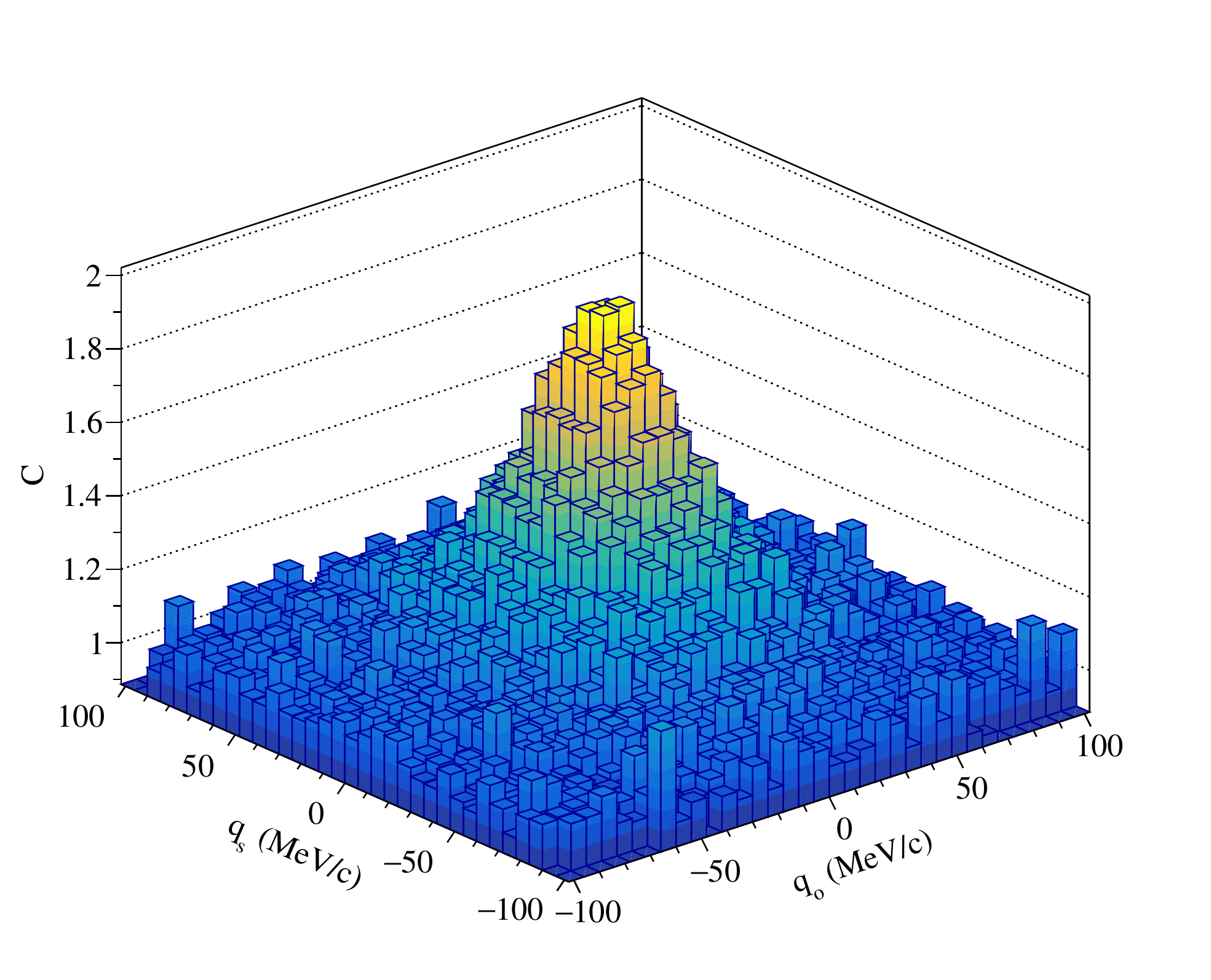}
	}
	\subfigure[]
	{
		\includegraphics[width=0.47\textwidth]{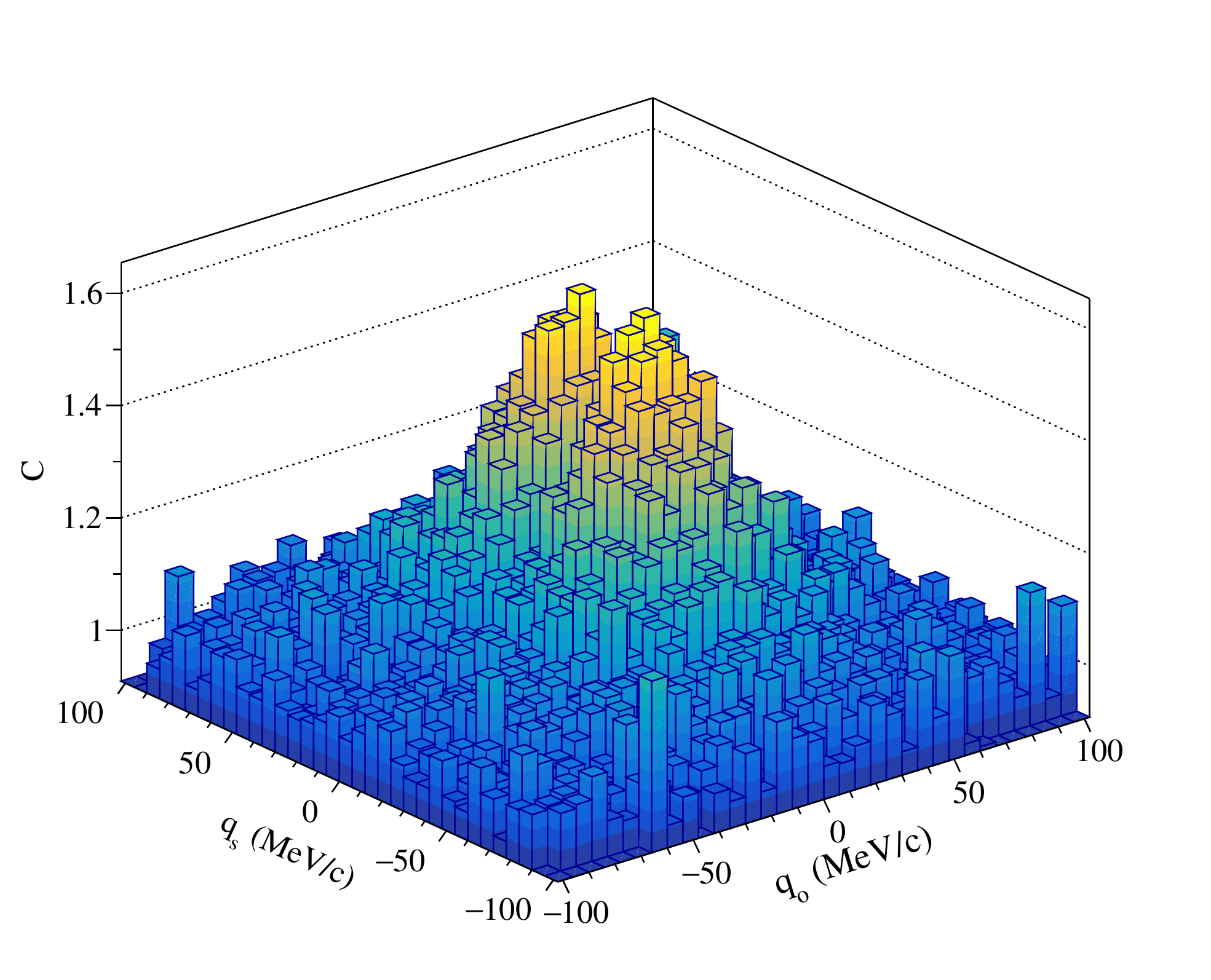}\label{fit_c_coul}
	}
	\caption{Correlation functions of $\pi^+$ in the $q_{\rm o}$ and $q_{\rm s}$ directions, generated by the string melting AMPT model for the Au+Au collisions at $\sqrt{S_{NN}}=$ 200 GeV. The $K_{\rm T}$ range is 325--375 MeV/c, and the $q_{\rm l}$ range is -3--3 MeV/c. Figure (a) neglects the Coulomb interaction, and figure (b) is the situation with the Coulomb interaction.}
	\label{fit_c}
\end{figure}

We only focus on the HBT radius $R_{\rm s}$, it is least affected by other physical factors and related to the transverse size of the source\cite{2011328}. In addition, the transverse momentum dependence of $R_{\rm s}$ with and without Coulomb interaction is shown in Figure~\ref{rs_p}. The obvious difference is the values of $R_{\rm s}$ calculated with the Coulomb interaction are smaller than those without the Coulomb interaction at small pair momentum. Besides, the strengths of $K_{\rm T}$ dependence of $R_{\rm s}$, especially for the lower collision energies, can barely be distinguished. And it is even worse for the situation with the Coulomb interaction. At low collision energies, there are similar collective flow velocities in each $K_{\rm T}$ bin, and it leads to similar single-particle momentum-space angle distributions\cite{Yang_2021}. And the single-particle momentum-space angle distributions can affect the HBT radii\cite{Yang_2020}, so the values of $R_{\rm s}$ are similar. For distinguish the strength of $K_{\rm T}$ dependence of $R_{\rm s}$, we introduced a parameter $b$, and it is in the fit function
\begin{equation} 
	R_{\rm s} = a K_{\rm T}^{b},
\end{equation}  
where parameter $a$ is just a common constant. The parameter $b$ can be used to describe the strength for the $K_{\rm T}$ dependence of $R_{\rm s}$, as shown in Figure~\ref{fig_e-b}.

\begin{figure}[!htb]
	\centering
	\subfigure[]
	{
		\includegraphics[width=0.37\textwidth]{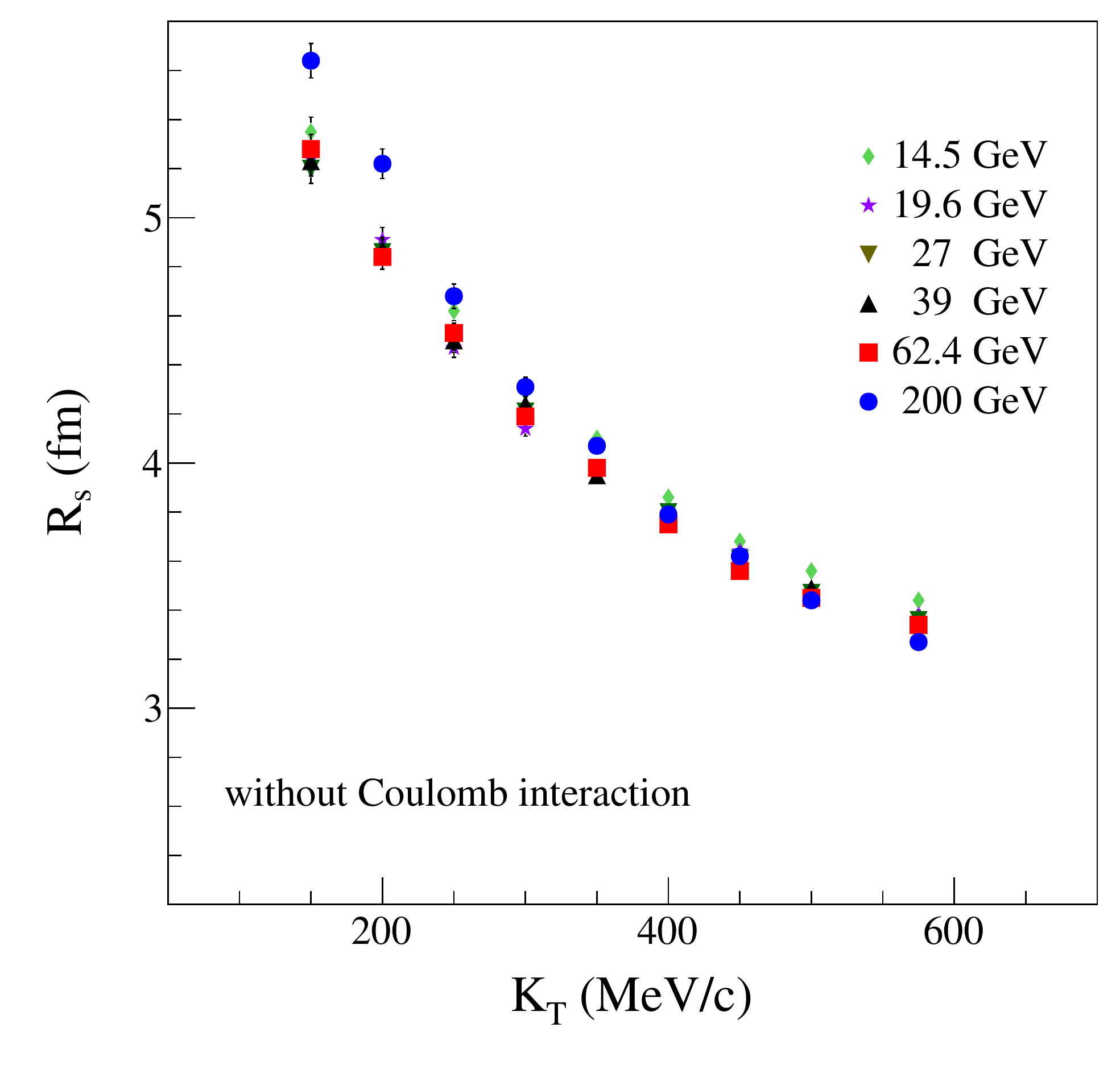}
	}
	\subfigure[]
	{
		\includegraphics[width=0.37\textwidth]{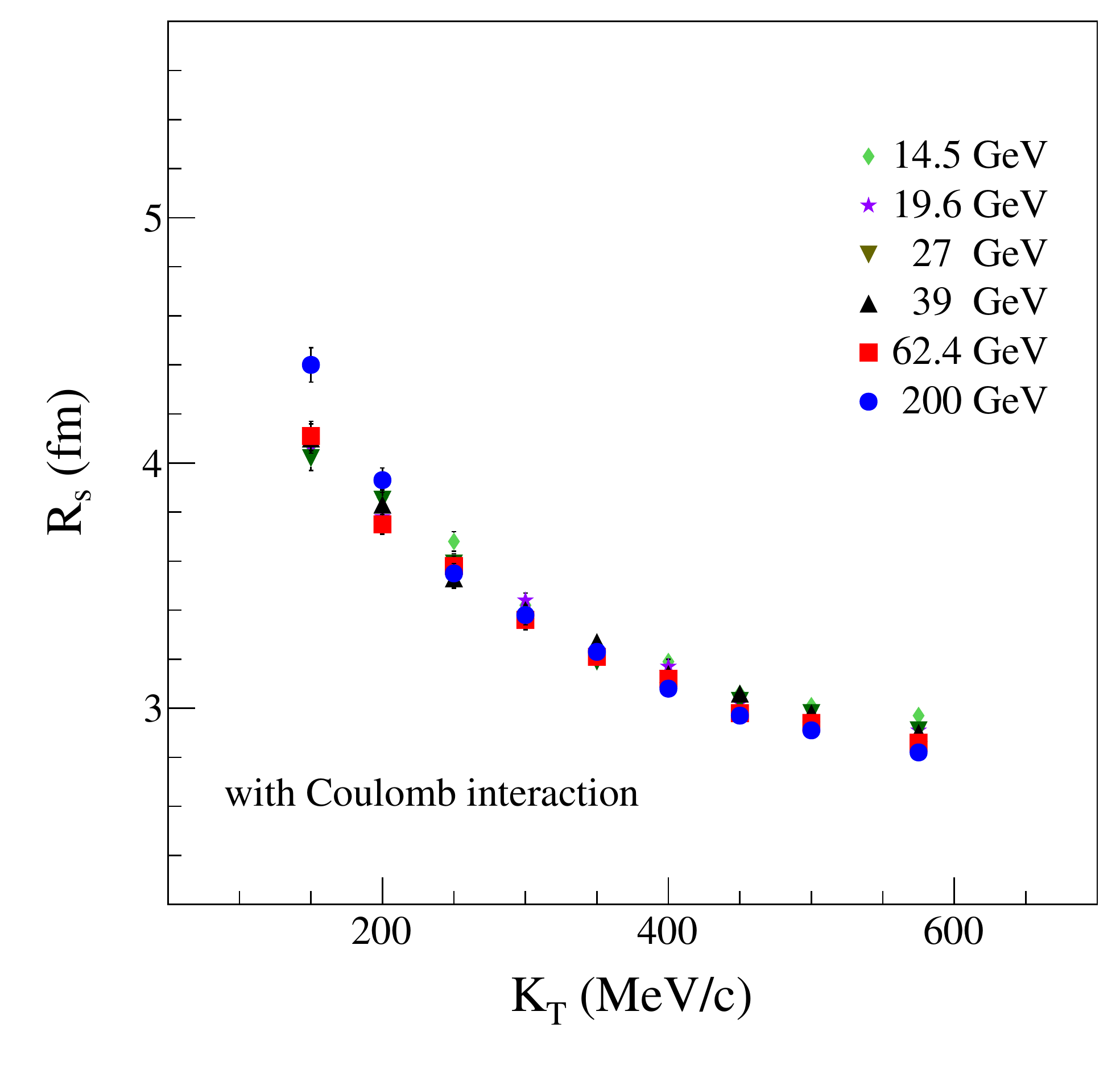}
	}
	\caption{Transverse momentum dependence of $R_{\rm s}$ in AMPT model.}
	\label{rs_p}
\end{figure}

\begin{figure}[!htb]
	\centering
	\includegraphics[scale=0.31]{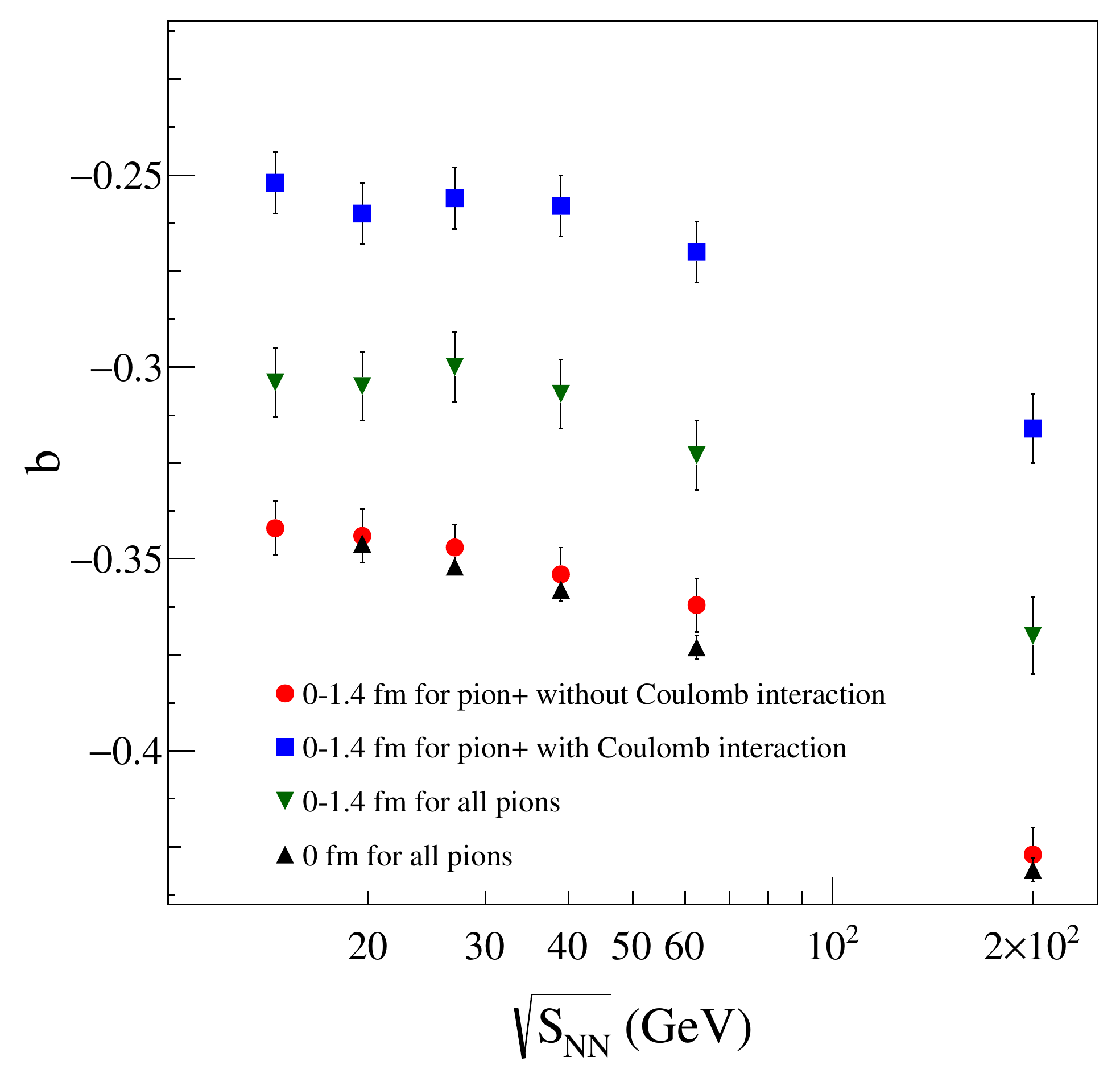}
	\caption{Collision energy dependence of parameter $b$.}
	\label{fig_e-b}
\end{figure} 

In Figure~\ref{fig_e-b}, All the $|b|$ values increase with increasing collision energies. It indicates that, at high collision energies, the changes in $R_{s}$ values are more intense with the transverse pair momenta. The black dots for all pions are taken from our last work, which treated all pions as one kind and generated with the 0 fm impact parameter. And comparing to the 0-1.4 fm for all pions, the $|b|$ values decrease with a higher impact parameter. The impact parameter can decrease the HBT radii\cite{PhysRevC.92.014904}, which indicates the impact parameter has different influences on $R_{\rm s}$ in different $K_{\rm T}$ sections. And compare $\pi^+$ without Coulomb interaction and all pions, shows that the kind of pions can also affect the strength of the $K_{\rm T}$ dependence of $R_{\rm s}$. The $|b|$ values are smaller with the Coulomb interaction, which means the Coulomb interaction inhibits the strength of the transverse momentum dependence of HBT radii. Furthermore, with increasing the collision energies, the differences of the $|b|$ values for $\pi^+$ between the two situations are also increased, which indicates this inhibition is stronger at high collision energies for the Coulomb interaction.

%%%%%%%%%%%%%%%%%%%%%%%%%%%%%%%%%%%%%%%%%%%%%%%%%%%%%%%%%%%%%%%%%%%%%%%%%%%%%%%%%%%%%%%%%%%%
\section{The single-particle space-momentum angle distribution}

In HBT research, the correlation function is often approximated by the on-shell momenta\cite{PhysRevC.49.442,PhysRevC.42.2646}. So the correlation can also be written as\cite{Chapman:1994ax}
\begin{equation}
C(\bm{q},\bm{K})\approx 1+\frac{|\int {\rm d}^4x S(x,K))
		e^{i \bm q \cdot \bm x}|^2}
		{|\int {\rm d}^{4}xS(x,K)|^2} \equiv 1+|<e^{i \bm q \cdot \bm x}>|^2,
\end{equation}
where $K_0=E_K=\sqrt{(m^2+|\bm K|^2)}$, $\bm q = \bm p_1-\bm p_2$, then we get 
\begin{equation}
e^{i\bm q \cdot \bm x}=\exp[i(\bm{p}_1 \cdot \bm{x}-\bm{p}_2 \cdot \bm{x})],
\end{equation}
for the single-particle, we have already defined the angle $\Delta\varphi$ which is between the momentum direction and space direction, then we define two more angles,
\begin{equation}
\cos(\Delta\alpha) \equiv \frac{\bm p_2 \cdot {\bm x}}{|\bm p_2||{\bm x}|},
\end{equation}
\begin{equation}
\cos(\Delta\beta) \equiv \frac{\bm p_1 \cdot \bm p_2}{|\bm p_1||\bm p_2|},
\end{equation}
where $\Delta\alpha$ is the angle between the momentum direction and radius direction for two particles, and $\Delta\beta$ is the angle between the momentum directions for two particles. And the smoothness assumption is $|\bm p_1| \approx |\bm p_2| \approx\frac{1}{2} |\bm K|$\cite{PhysRevC.56.1095}, then that
\begin{equation}
e^{i\bm q \cdot \bm x}\approx\exp\left[\frac{i}{2}\left(|\bm K||x|\cos(\Delta\varphi) - |\bm K||x|\cos(\Delta\alpha)\right)\right],
\end{equation}
the $\Delta\alpha$ angle is not independent, its values are related to the $\Delta\beta$ and the $\Delta\varphi$, $\Delta\alpha=f(\Delta\varphi,\Delta\beta)$, so the correlation function can be written as
\begin{equation}
C(\bm K, \Delta\varphi, \Delta\beta) \approx 1+\left|\left<\exp\left[ \frac{i}{2}|\bm K||\bm x|[\cos(\Delta\varphi)-\cos(f(\Delta\varphi,\Delta\beta))] \right]\right>\right|^2.
\end{equation}

When we calculate the HBT correlation function, we need to limit the values of momentum difference, $|\bm q|<q_{\rm {max}}$. If we choose the pair particles in one transverse momentum section, and because we focus on the mid-rapidity range, the pair momenta are also screened, $k_{\rm {min}}<|\bm K|<k_{\rm {max}}$. In the same time, the angle $\Delta\beta$ is also limited, $\cos(\Delta\beta)>\frac{2k_{\rm {min}}^2}{k_{\rm {min}}^2+q_{\rm {max}}^2}-1$. So the $\Delta\beta$ values are related to the $|\bm K|$ values and the interaction between the particles. Therefore, the single-particle space-momentum angle distribution can directly affect the correlation function. And our research is focused on the transverse plane, so we discuss the connection between the $\Delta\theta$ distribution and $R_{\rm s}$.

For a source, the collective flow is created by the expansion, and the transverse flow velocities are different at different layers of the source, which leads to different normalized $\Delta\theta$ distributions, as shown in Figure~\ref{fig_rt_cos}. 
\begin{figure}[!htb]
	\centering
	\includegraphics[scale=0.23]{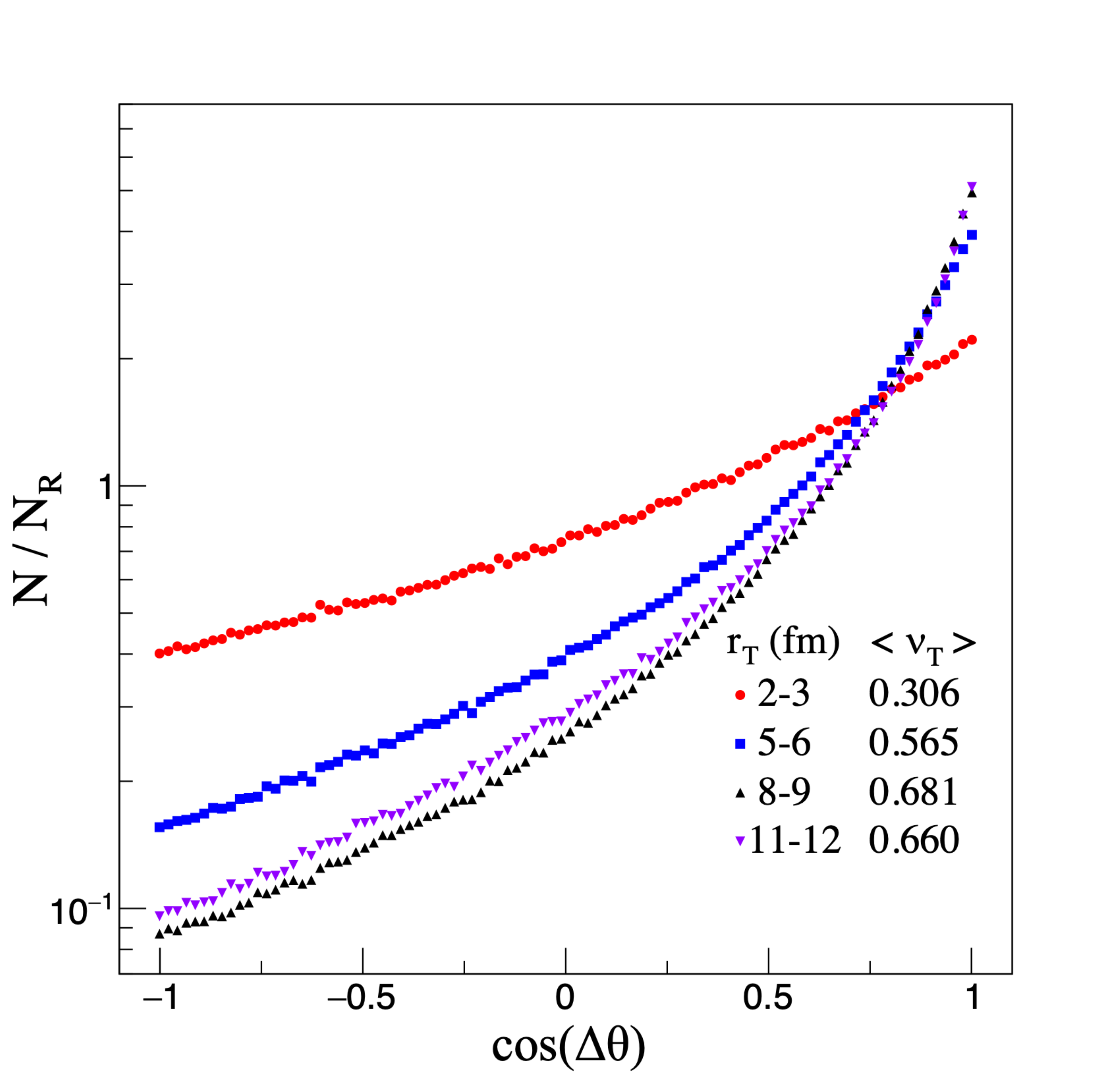}
	\caption{The normalized $\Delta\theta$ distributions in different transverse radii, $\pi^{+}$ are generated by the string melting AMPT model for the Au+Au collisions at $\sqrt{S_{NN}}=$ 62.4 GeV/c.}
	\label{fig_rt_cos}
\end{figure}

In Figure~\ref{fig_rt_cos}, the average transverse flow velocity can be calculated by
\begin{equation}
<v_{\rm T}> = \left< \frac{\bm p_{\rm T} \cdot \bm r_{\rm T}}{E|\bm r_{\rm T}|} \right>. 
\end{equation} 
$\rm N$ and $\rm N_R$ are the particle numbers in each bin, while $\rm N_R$ is obtained with the random $\bm p$ and $\bm r$ particles. The normalization process is using the $\cos(\Delta\theta)$ distribution divided by the random $\cos(\Delta\theta)$ distribution, and we let $\sum \rm N =\sum \rm N_R$. The particles with lower transverse momenta are closer to $N/N_{\rm R} = 1$, this phenomenon indicates the source is approaching a random freeze-out source. While with higher flow velocities, the distributions are closer to $\cos(\Delta\theta)=1$, and it means the particles tend to freeze out along the radius direction. And for the particles that are located at 8-9 and 11-12 fm, their $<\nu_{\rm T}>$ values are closer, so they have similar normalized $\Delta\theta$ distributions.

Furthermore, for the particles with different transverse momenta $p_{\rm T}$, they correspond to the parts of the whole source. And the particles with higher $p_{\rm T}$, the part sources have bigger collective flow, as shown in Figure~\ref{fig_vt_rt}. The $<\nu_{\rm T}>$ increase with $r_{\rm T}=\sqrt{x^2+y^2}$ inside of the source, and at the outside of the source, $<\nu_{\rm T}>$ almost becomes a constant. So the sizes of part sources are changing with the $p_{\rm T}$.

\begin{figure}[phtb]
	\centering
	\includegraphics[scale=0.3]{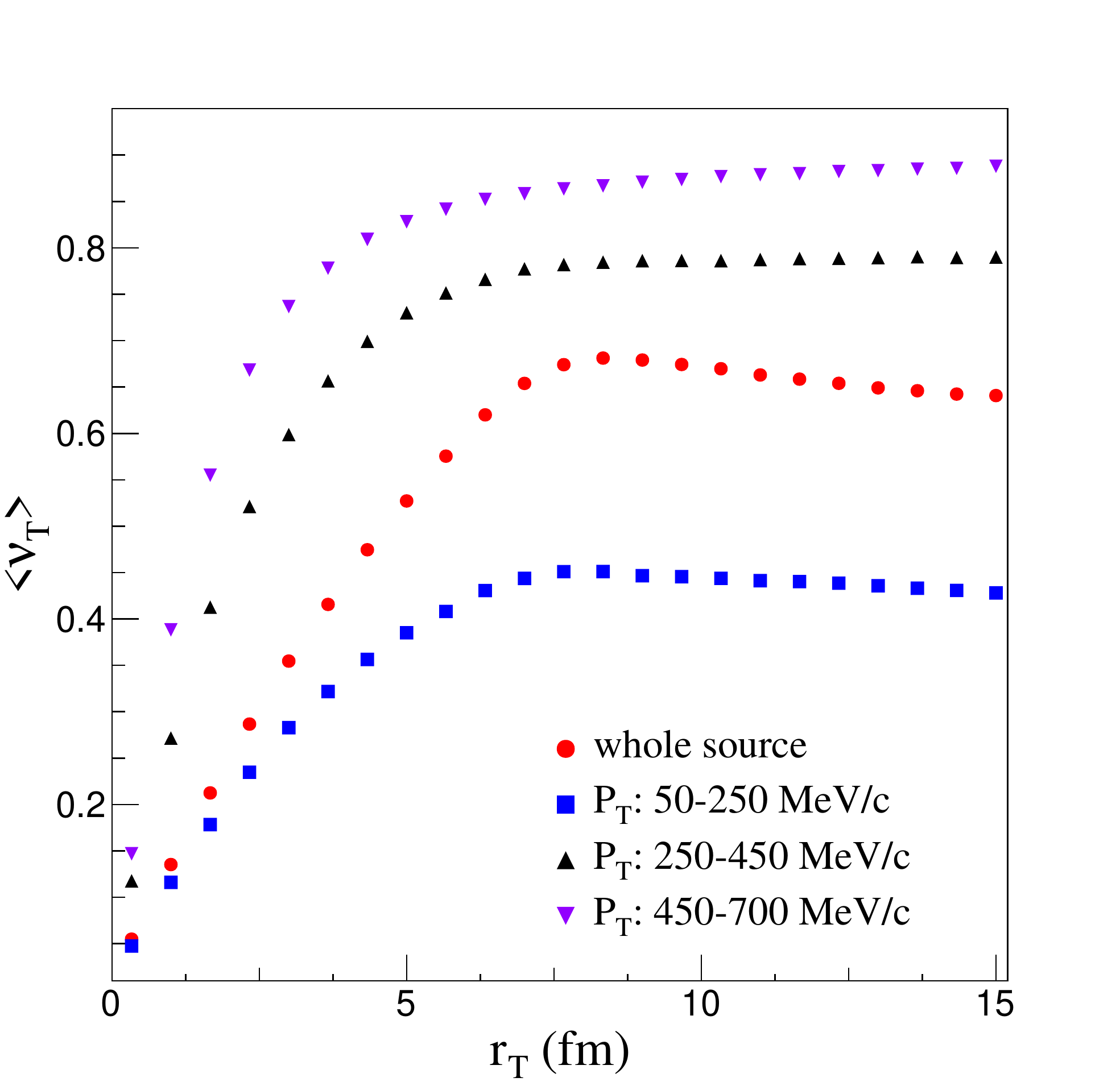}
	\caption{The transverse flow velocities for the transverse radii in different transverse momenta, $\pi^{+}$ are generated by the string melting AMPT model for the Au+Au collisions at $\sqrt{S_{NN}}=$ 62.4 GeV/c.}
	\label{fig_vt_rt}
\end{figure}

Before particle freezing out from the source, the system goes through a series of processes, containing parton production and interaction, hadronization, and hadron cascade. And the expansion of the source will be reflected in the collective flow of the freeze-out particles. The collective flow is different at different locations in the source, meanwhile, the $\Delta\theta$ is related to the flow, different $\Delta\theta$ have different freeze-out sources, as shown in Figure~\ref{fig_cos}. 

\begin{figure}[!htb]
	\centering
	\subfigure[$-1<\cos(\Delta\theta)<0.6$]
	{
		\includegraphics[width=0.365\textwidth]{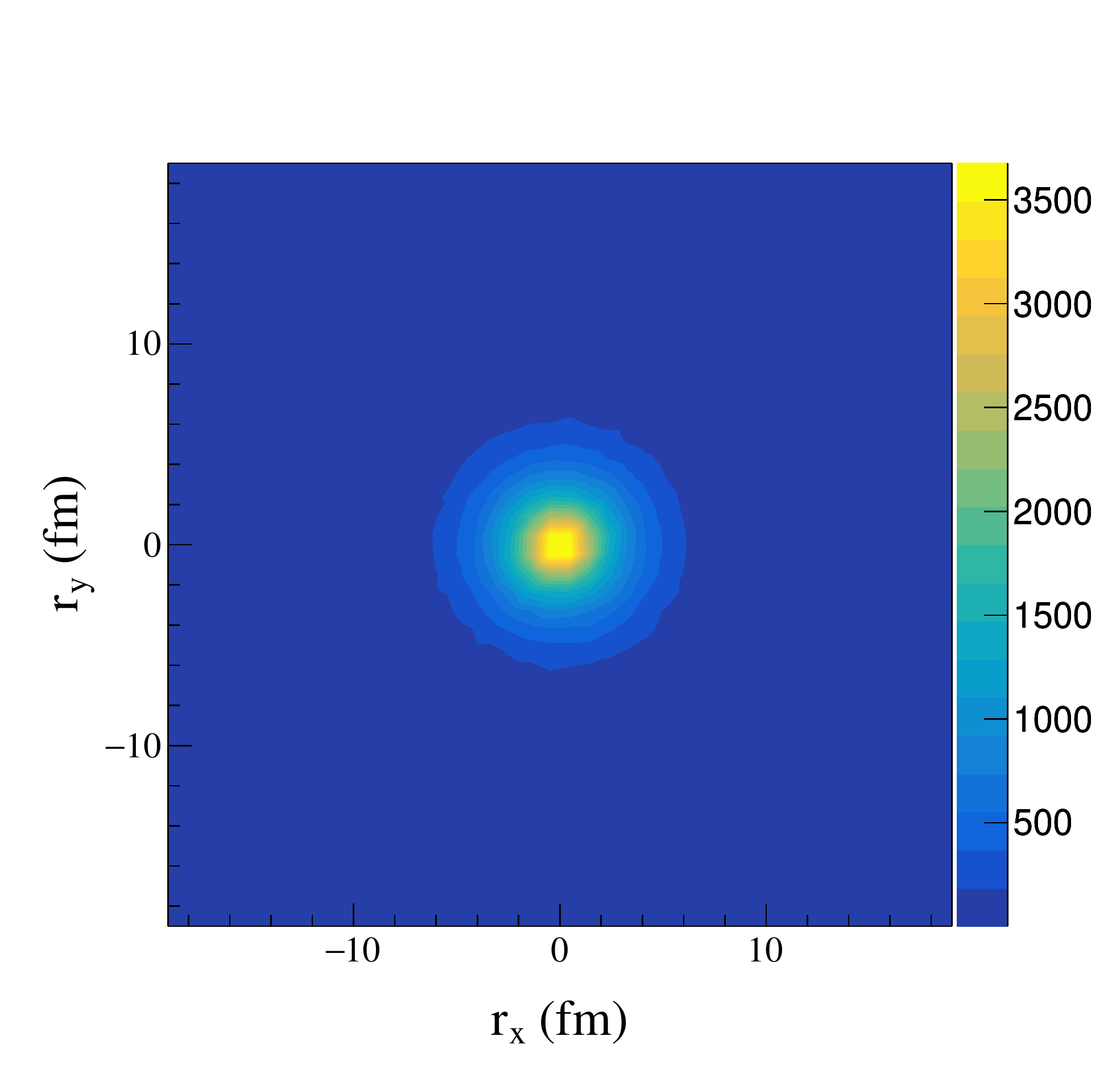}\label{fig_cos_rev}
	}
	\subfigure[$-0.2<\cos(\Delta\theta)<0.2$]
	{
		\includegraphics[width=0.365\textwidth]{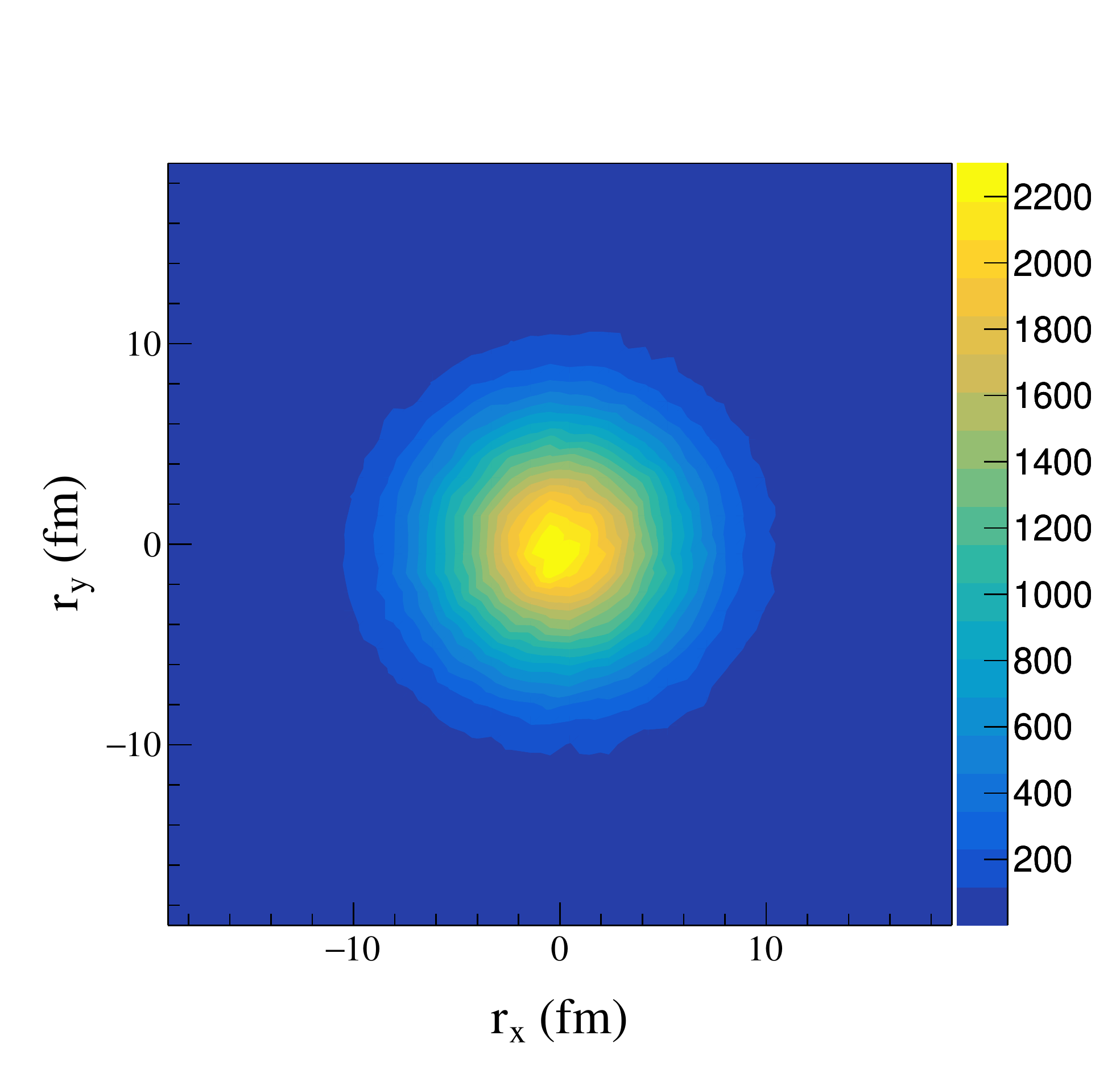}\label{fig_cos_per}
	}
	\subfigure[$0.6<\cos(\Delta\theta)<1$]
	{
		\includegraphics[width=0.365\textwidth]{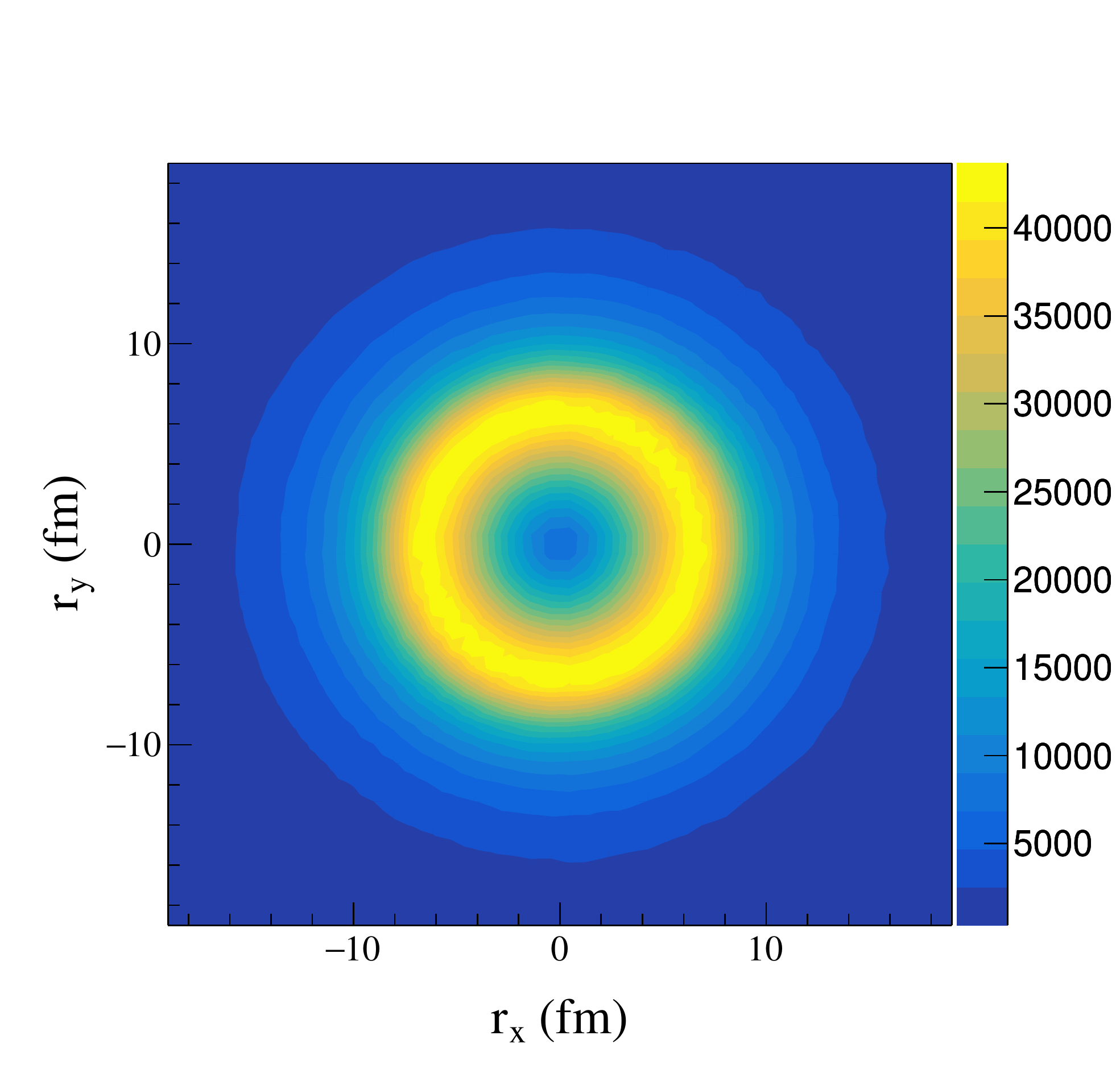}\label{fig_cos_alo}
	}
	\caption{The freeze-out $\pi^+$ sources in different $\cos(\Delta\theta)$ sections. The particles are generated by the string melting AMPT model for the Au+Au collisions at $\sqrt{S_{NN}}=$ 62.4 GeV, and the $p_{\rm T}$ range is 250--450 MeV/c.}
	\label{fig_cos}
\end{figure}

For the $\pi^+$ that have large $\Delta\theta$ angles, their freeze-out directions are reversed to the space directions. And they have a small freeze-out source, which indicates they are almost freezing out from the center of the whole $\pi^+$ source. Those $\pi^+$ freeze-out directions are perpendicular to the space directions and have a bigger source, as shown in Figure~\ref{fig_cos_per}. While in Figure~\ref{fig_cos_alo}, for the $\pi^+$ whose momentum directions along to the space directions, their freeze-out source has a ring shape, and has the biggest number density. So in this $p_{\rm T}$ section, most of the $\pi^+$ freeze out from the shell. And the closer to the center of the source, the higher the probability of producing large $\Delta\theta$ angle particles. For we limited the differences of the pair particles $q_{\rm max}$, the normalized $\cos(\Delta\theta)$ distribution in $K_{\rm {Tmin}}<K_{\rm T}<K_{\rm {Tmax}}$ section, it only contains $K_{\rm Tmin}-q_{\rm max}/2 <p_{\rm T}< K_{\rm Tmax}+q_{\rm max}/2$ particles. Besides, different $\cos(\Delta\theta)$ in the $K_{\rm T}$ section has different partial sources, they are similar to Figure~\ref{fig_cos}. The normalized $\cos(\Delta\theta)$ distribution corresponds to the superposition of these partial sources. Therefore, the distribution is related to the HBT radius $R_{\rm s}$.

If there is no correlation between the space and momentum, it will break the relation between $\Delta\theta$ and the source, and the differences in the normalized $\cos(\Delta\theta)$ distributions will disappear. Then the $\Delta\theta$ angle will be completely random, as shown in Figure~\ref{fig_no_phi}. The normalized "no x-p correlation" distribution becomes a line and is the same in other $K_{\rm T}$ sections.
\begin{figure}[!htb]
	\centering
	\includegraphics[scale=0.21]{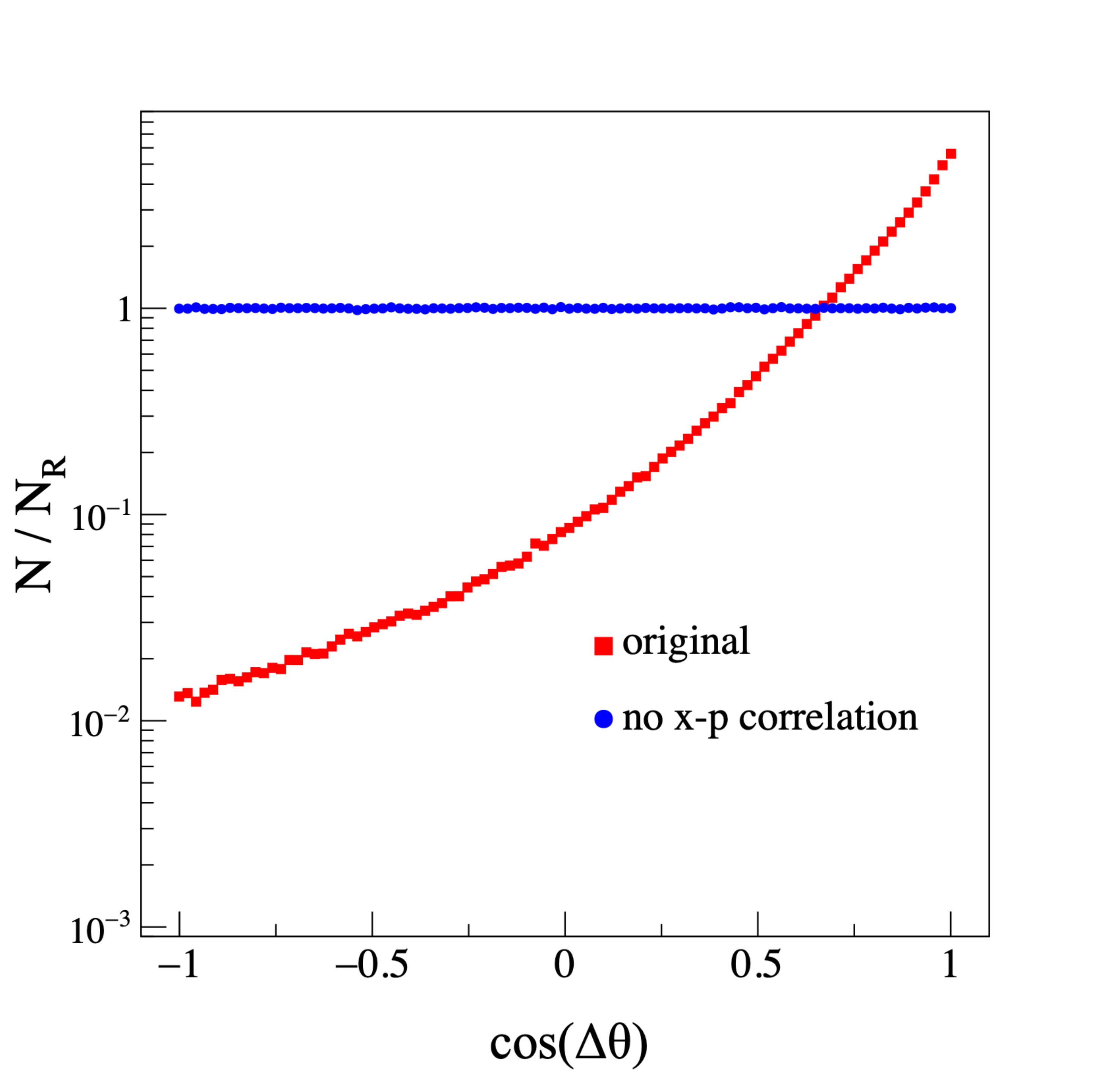}
	\caption{The original distribution and the no x-p correlation distribution at $\sqrt{S_{NN}}=$ 62.4 GeV, and the $K_{\rm T}$ range is 325--375 MeV/c.}
	\label{fig_no_phi}
\end{figure}

In Figure~\ref{fig_no_phi}, we disrupt the space and the momentum correlation in each collision event. For example, original data are $x_{1}p_{1}$ and $x_{2}p_{2}$, and after the disruption, they are $x_{1}p_{2}$ and $x_{2}p_{1}$. This is a rough but effective method, and it can change the correlation between the space and the momentum, which means the particles can freeze out from the source in any direction. Then we calculate the HBT radius $R_{\rm s}$, as shown in Figure~\ref{fig_no_rs}. The results show that the phenomenon of transverse dependence of $R_{\rm s}$ for "no x-p correlation" is almost disappeared. The $R_{\rm s}$ values for "no x-p correlation" are closer to each other, it is because they have the same normalized $\cos(\Delta\theta)$ distribution. And this method destroys the physical process, there will be fluctuations in calculating the correlation function, which leads to the emergence of the big error bars.

\begin{figure}[!htb]
	\centering
	\includegraphics[scale=0.33]{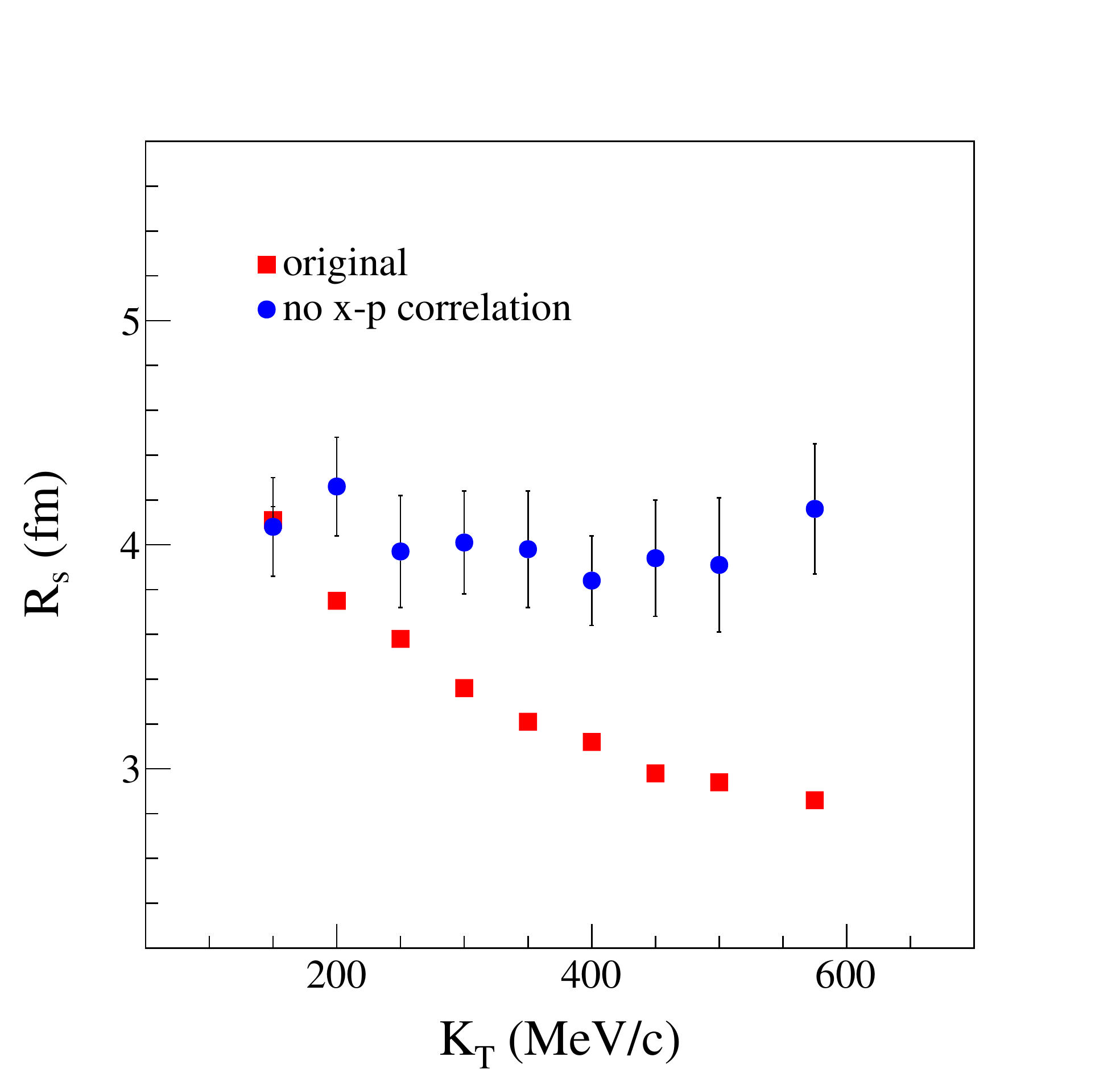}
	\caption{The original $R_{\rm s}$ and no x-p correlation $R_{\rm s}$ at $\sqrt{S_{NN}}=$ 62.4 GeV.}
	\label{fig_no_rs}
\end{figure}

Moreover, we can build a numerical connection between the normalized $\cos(\Delta\theta)$ distribution and HBT radius $R_{\rm s}$. The AMPT model does not contain the Coulomb interaction at the stage of the hadron cascade\cite{PhysRevC.72.064901}, so we discuss the Coulomb interaction only for the HBT analysis. And the two conditions are used the same string melting AMPT data, leading to the same normalized $\cos(\Delta\theta)$ distribution for different $K_{\rm T}$ regions. In order to describe the normalized $\cos(\Delta\theta)$ distributions, we introduce the fit function
\begin{equation} 
	f=0.002\exp\bigg\{ c_{1}\exp\Big[c_{2}\cos(\Delta\theta) \Big] \bigg\}.
\end{equation}
$c_1$ and $c_2$ are the fit parameters. $c_1$ is influenced by the proportion of particles whose $\cos(\Delta\theta)=0$, and $c_2$ is influenced by the strength of the distribution approaching $\cos(\Delta\theta)=1$. The value of 0.002 is settled by us to get good fitting results, and this value is different than our last work because the normalized $\cos(\Delta\theta)$ distributions have changed, as shown in Figure~\ref{fig_costheta}. We can see in two $K_{\rm T}$ sections, $\pi^+$ and all pions with 0-1.4 fm impact parameter have similar distributions, while the impact parameter has a big influence on the distributions. So the impact parameter plays an important role in the single-particle space-momentum angle correlation. For we change this settled value, it leads to the changing of the fitting results in the subsequent analysis. The parameters $c_1$ and $c_2$ are changing with the $K_{\rm T}$, so we can use the fitting functions to describe their changing pattern. The fit functions are
\begin{eqnarray} 
	c_{1}=k_{1} \exp\Big[-6\times(\frac{K_{\rm T}}{1000})^{2} \Big]+j_{1},\\
	c_{2}=k_{2} \exp\Big[-4.5\times(\frac{K_{\rm T}}{1000})^{2} \Big]+j_{2},
\end{eqnarray} 
where -6 and -4.5 were chosen by us to obtain good fitting results, k and j are fit parameters. $k_1$ and $j_1$ are related to $c_1$. $k_1$ is influenced by the changing range of the proportion of pions with $\cos(\Delta\theta)=0$, and $j_1$ is influenced by this lowest proportion. Besides, $k_2$ and $j_2$ are related to $c_2$, $k_2$ is influenced by the changing range of the strength of the distribution approaching $\cos(\Delta\theta)=1$, and $j2$ is influenced by its highest strength.

\begin{figure}[!htb]
	\centering
	\includegraphics[scale=0.22]{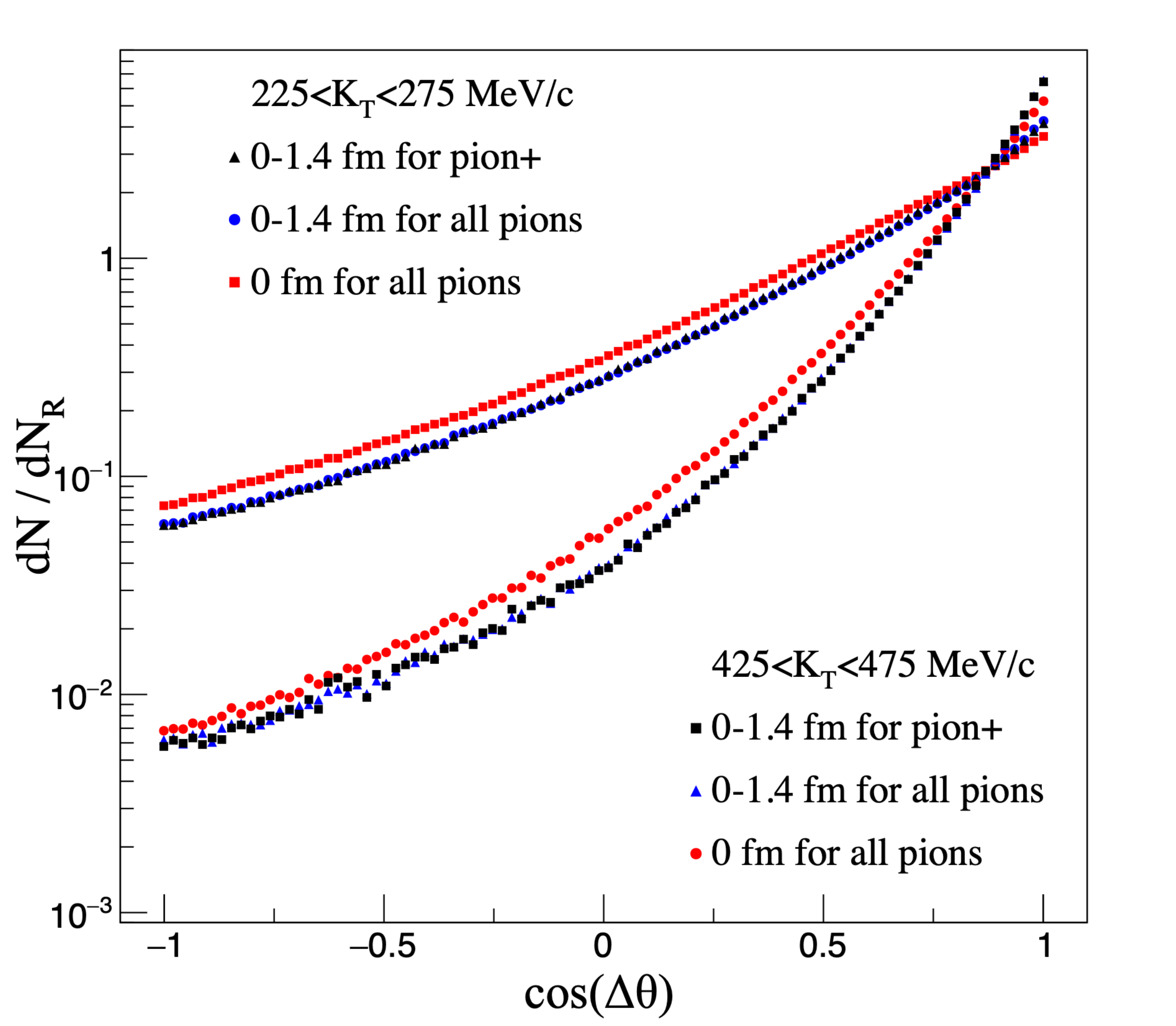}
	\caption{The normalized $\Delta\theta$ distributions, generated by the string melting AMPT model for the Au+Au collisions at $\sqrt{S_{NN}}=$ 39 GeV/c.}
	\label{fig_costheta}
\end{figure} 

The collision energies not only have influences on the $K_{T}$ dependence of $R_{s}$, but also the $K_{T}$ dependence of normalized $\cos(\Delta\theta)$ distributions. Then we can plot the parameter $b$, as the functions of the parameters $k$ and $j$, as shown in Figure~\ref{fig_b_j_k}. The changing patterns indicate parameter b has an extremum, we set $b_1=-0.2$ for the situation without the Coulomb interaction, and $b_2=-0.3$ for considering the Coulomb interaction. We use the functions to fit these patterns, they are
\begin{eqnarray}
	b(k_m) = \mu_{m1} |k_m| ^{\mu_{m2}}+b_n,\label{bk1}    \\
	b(j_m) = \nu_{m1} |j_m| ^{\nu_{m2}}+b_n,\label{bj2}		
\end{eqnarray}
where $m=$1 or 2, it is for distinguishing the parameters of $c_1$ or $c_2$. Also $n=$1 or 2, then 1 is for the situation without the Coulomb interaction, and 2 means the Coulomb interaction. $\mu$ and $\nu$ are fit parameters, and the fitting results are shown in Table~\ref{muandnu}. The results show that the Coulomb interaction has an influence on these fit parameters, it can decrease $|\mu_{m1}|$ and $|\nu_{m1}|$, and increase $|\mu_{m2}|$ and $|\nu_{m2}|$. It is caused by the changing of the differences of $b$ values. And the differences in $b$ values between the two situations are increasing with increasing the collision energies. Compared with our previous work, the changing patterns are different, so these numerical connections are also related to the kinds of particles.
\begin{figure}[!htb]
	\centering
	\subfigure[]
	{
		\label{k1_b}
		\includegraphics[width=0.33\textwidth]{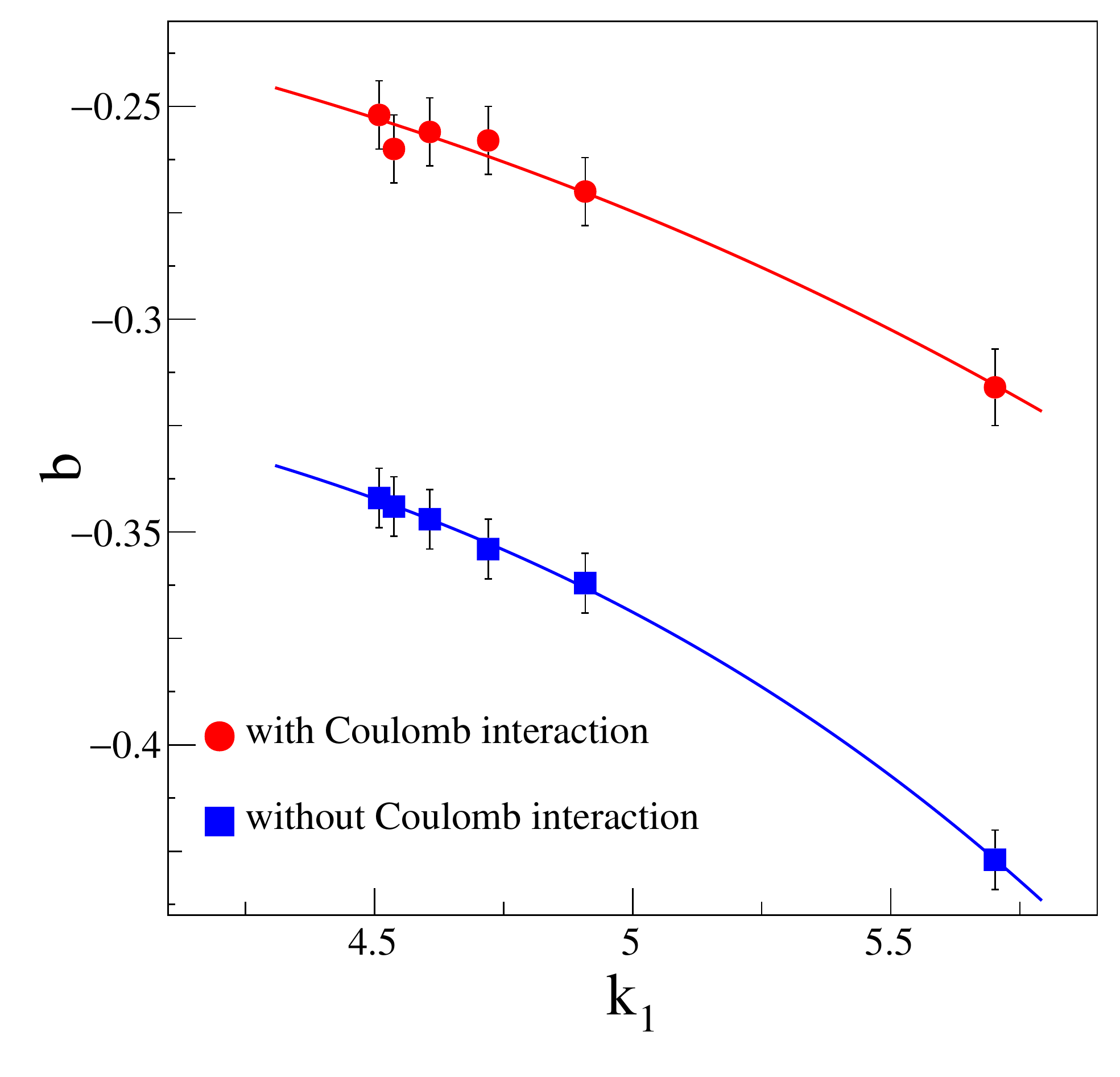}
	}
	\subfigure[]
	{
		\label{j1_b} 
		\includegraphics[width=0.33\textwidth]{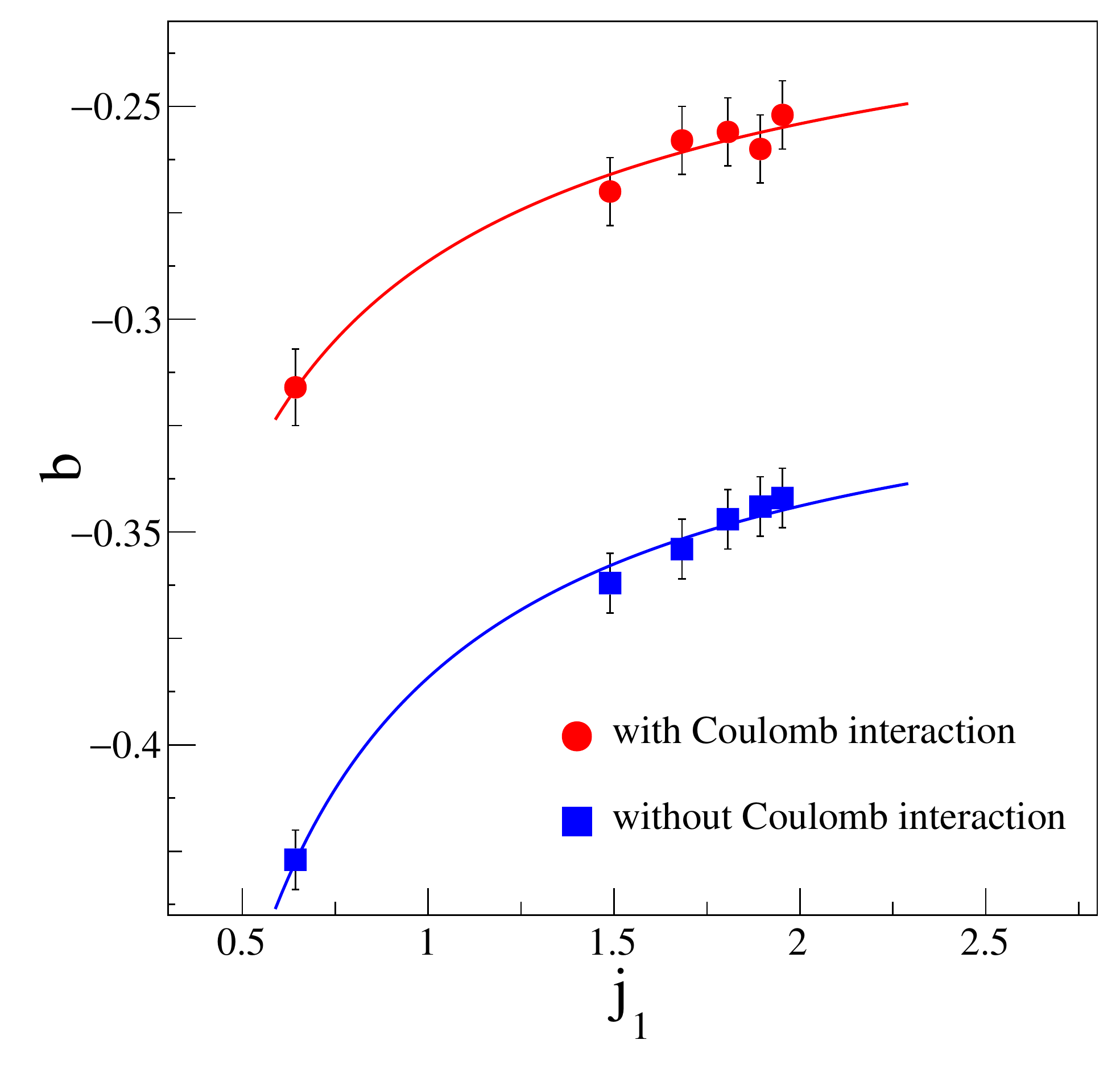}
	}
	
	\subfigure[]
	{
		\label{k2_b}
		\includegraphics[width=0.33\textwidth]{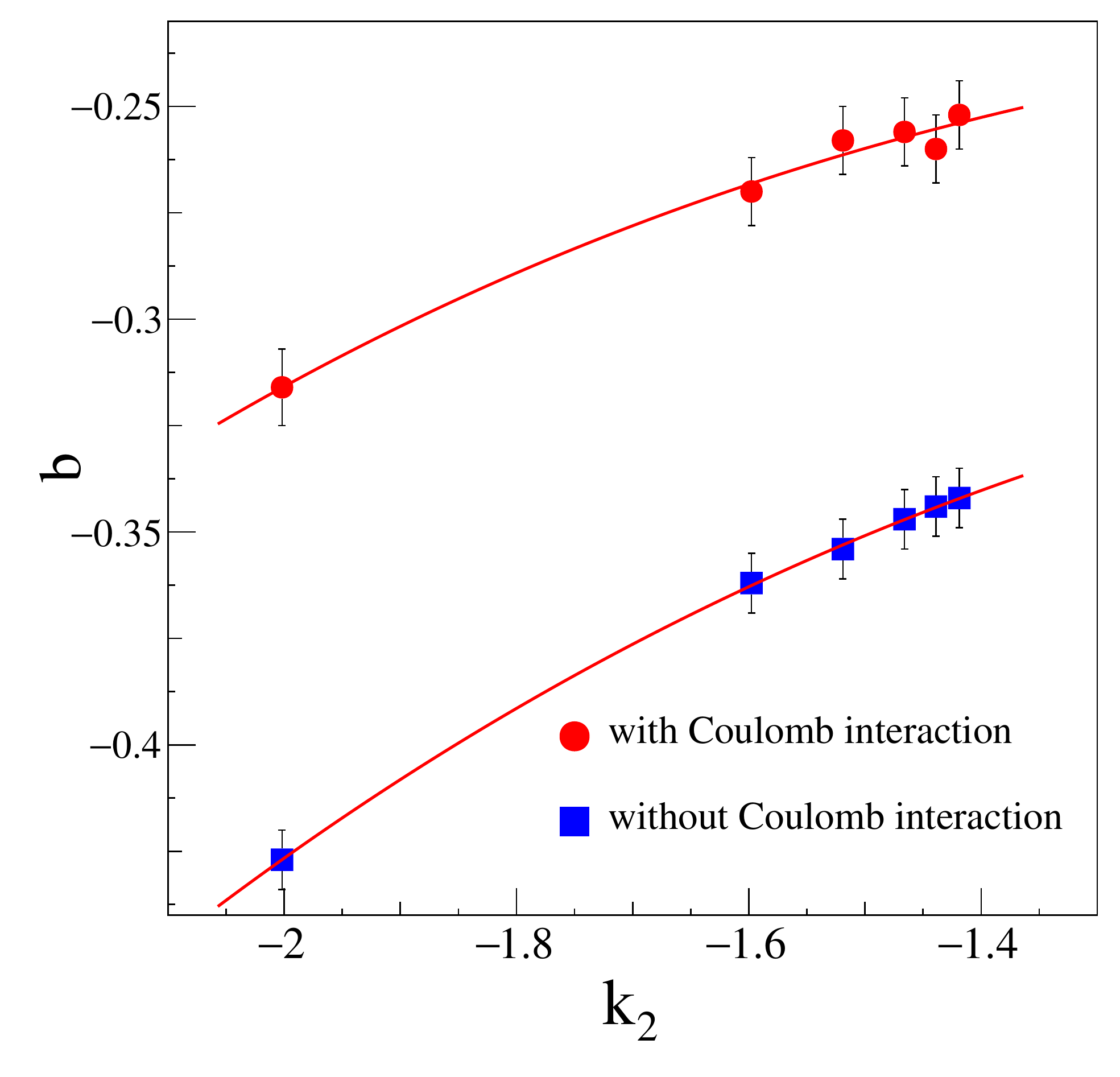}
	}
	\subfigure[]
	{
		\label{j2_b} 
		\includegraphics[width=0.33\textwidth]{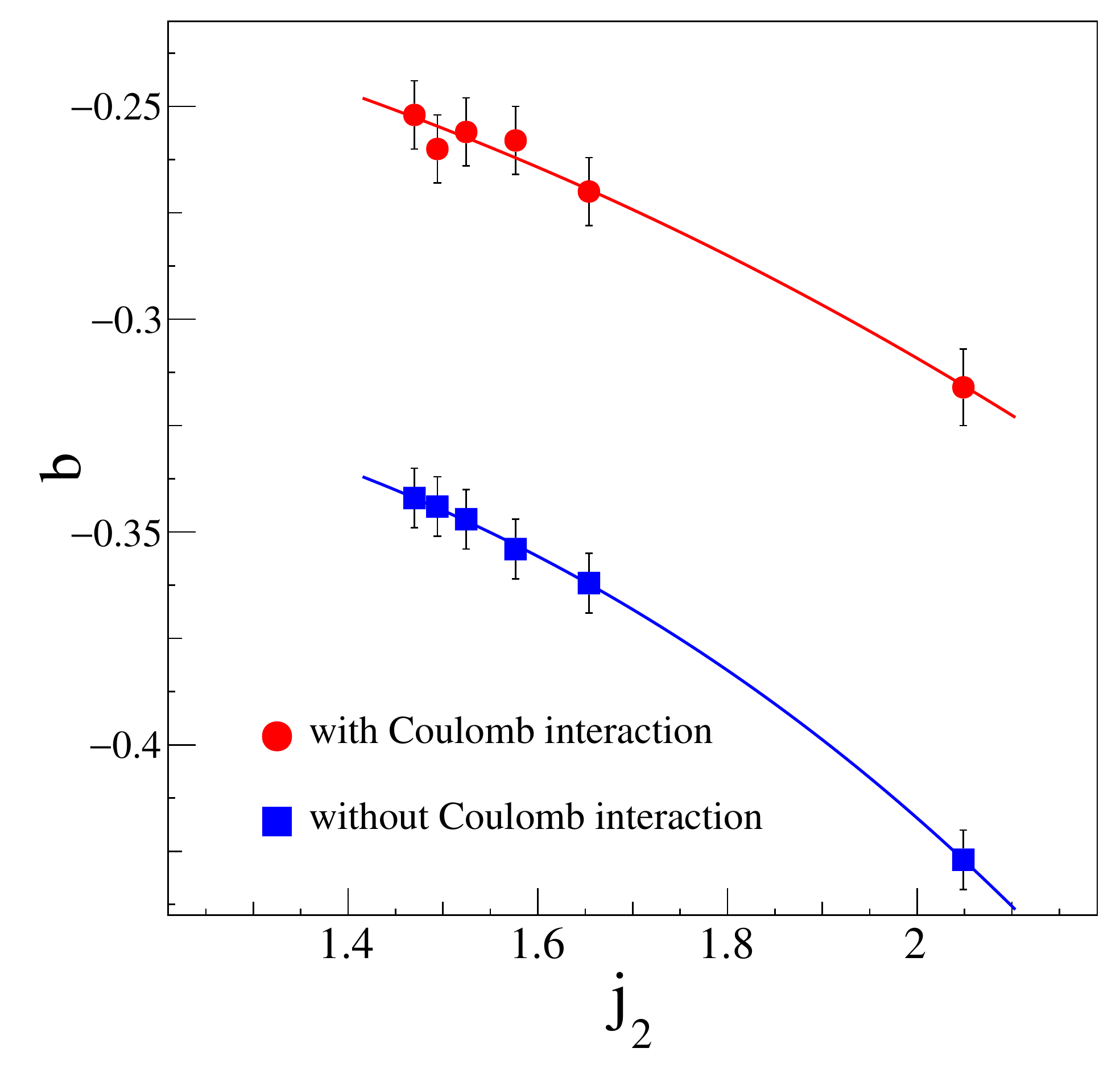}
	}
	\caption{The numerical connections between the strength of the transverse momentum dependence of $R_{\rm s}$ and the parameters related to the normalized $\cos(\Delta \theta)$ distribution, the lines are fit lines.}
	\label{fig_b_j_k}
\end{figure}

\begin{table}[!htb]
\begin{indented}
\lineup
\item[]
	\caption[table]{Fit results for parameters $\mu$ and $\nu$}\label{muandnu}
	\begin{tabular}{@{}*{4}{cccc}}
		\br
		{}& {} & $n=1$ & $n=2$\\
		\mr
		\multirow{4}{1 cm}{m=1}& $\mu_{11}$ & $-0.0004\pm0.0003$ & $-0.00004\pm0.00003$	\\
							   & $\mu_{12}$ & 	  $3.3\pm0.5$    & 		$4.7\pm0.4$	    \\
							   & $\nu_{11}$ &  $-0.086\pm0.004$  &   $-0.084\pm0.003$	\\
							   & $\nu_{12}$ &   $-0.68\pm0.09$   &    $-0.94\pm0.08$	\\

		\multirow{4}{1 cm}{m=2}& $\mu_{21}$ &  $-0.024\pm0.003$  &   $-0.014\pm0.003$	\\
							   & $\mu_{22}$ & 	  $2.3\pm0.3$    & 		$3.2\pm0.3$	    \\
							   & $\nu_{21}$ &  $-0.021\pm0.004$  &   $-0.012\pm0.002$	\\
							   & $\nu_{22}$ &     $2.4\pm0.3$    &      $3.3\pm0.3$	    \\
		\br
	\end{tabular}
\end{indented}
\end{table}  

With these fit functions and parameters, the numerical connections between the normalized $\cos(\Delta\theta)$ distribution and the strength of the transverse momentum dependence of $R_{\rm s}$ have been built. And comparing with our previous work, we found that the fit parameters are related to the impact parameter, the kinds of pions, and the Coulomb interaction. We improve the fit parameters and it makes our numerical correlation applied to the experiment closer. With further research, the experiment data can be directly used to study the single-particle space-momentum correlation.
%%%%%%%%%%%%%%%%%%%%%%%%%%%%%%%%%%%%%%%%%%%%%%%%%%%%%%%%%%%%%%%%%%%%%%%%%%%%%%%%%%%%%%%%%%%%
\section*{Conclusions}

With the string melting AMPT model, we calculate the HBT radius $R_{\rm s}$ for $\pi^+$ in two situations, with and without Coulomb interaction. The results indicate the Coulomb interaction can decrease the $R_{\rm s}$ values and inhibits the $K_{\rm T}$ dependence of $R_{\rm }$. And we compare the strength of $K_{\rm T}$ dependence of $R_{\rm }$ in two impact parameters, and found that the impact parameters can also inhibit $K_{\rm T}$ dependence of $R_{\rm }$. Then we show the transverse flow can affect the normalized $\cos(\Delta\theta)$ distributions, and the flow changes with the location of the source, which leads to the particles with bigger $\Delta\theta$ angles tending to freeze out from the inner of the source. We also present the normalized $\cos(\Delta\theta)$ distributions can influence the $K_{\rm T}$ dependence of $R_{\rm s}$ by disrupting the space and the momentum correlation. While the inhibition of the transverse momentum dependence of $R_{\rm s}$ caused by Coulomb interaction leads to the changing of the fit results. And different impact parameters and particles have different numerical connections. Moreover, with these numerical connections, we can get more information about the final stage of the Au+Au collision at the freeze-out time by the HBT analysis. And with more collision energies, more impact parameters, and more accurate fits, the numerical connection can be improved. There are anomalies near the CEP, it may change this connection, so the improvement connection may be a probe to detect the CEP. And the Coulomb interaction can weaken the strength of the transverse momentum dependence of $R_{\rm s}$, it may also weaken the sensitivity of this probe.

%%%%%%%%%%%%%%%%%%%%%%%%%%%%%%%%%%%%%%%%%%%%%%%%%%%%%%%%%%%%%%%%%%%%%%%%%%%%%%%%%%%%%%%%%%%%
\section*{References}
\bibliography{references}

\end{document}